\documentclass[12pt]{iopart}

\usepackage{color}
\usepackage{epsfig}
\begin{document}

\newcommand{\n}{\nonumber}
\newcommand{\be}{\begin{equation}}
\newcommand{\ee}{\end{equation}}
\newcommand{\bea}{\begin{eqnarray}}
\newcommand{\eea}{\end{eqnarray}}
\newcommand{\lam}{\lambda}

\title[Prepotential Approach: an overview]{Prepotential Approach: a unified approach to exactly, quasi-exactly, and rationally extended solvable quantal systems}

\author{Choon-Lin Ho
 \footnote{email: hcl@mail.tku.edu.tw}}
\address{Department of Physics, Tamkang University,
Tamsui 25137, Taiwan (R.O.C.)}


\begin{abstract}

We give a brief overview of a simple and unified way, called the prepotential approach,  to treat  both exact
and quasi-exact solvabilities of the one-dimensional Schr\"odinger
equation.  It is based on the prepotential together with Bethe
ansatz equations.
Unlike the the supersymmetric method for the exactly-solvable systems and 
the Lie-algebraic approach for the quasi-exactly solvable problems,  this approach does not require
any knowledge of the underlying symmetry of the system.  It treats
both quasi-exact and exact solvabilities on the same footing.
  In this approach
the system is completely defined by the choice of two polynomials and a set of Bethe ansatz equations. 
The potential, the change of variables as well as the eigenfunctions and eigenvalues are determined
in the same process. We
illustrate the  approach by  several paradigmatic examples of Hermitian and
non-Hermitian Hamiltonians with real energies.  Hermitian systems with complex energies, called the quasinormal modes, are also presented. Extension of the approach to the newly discovered rationally extended models is briefly discussed.

\end{abstract}




\section{Introduction}

In this review I would like to give an overview of  a constructive approach to
both exact and quasi-exact solvable one-dimensional Schr\"odinger
equations with Hermitian and non-Hermitian Hamiltonians, 
and rationally extended solvable quantal systems \cite{Ho1,Ho2,Ho3,Ho4,Ho5,Ho6}.

Everyone knows what exact solvability means.  In fact, we learned the principles of quantum
mechanics mainly through several well-known exactly solvable (ES) models,
such as the infinite square well, the harmonic oscillator, the
hydrogen atom, etc. But in the real world, ES systems
are rather scanty. Most systems we encounter are non-solvable,
and we have to resort to various approximation schemes or numerical methods to solve them.

It is therefore of great interest to witness some new developments in recent years that have deepened 
our understanding of the solvability of the Schr\"odinger equation, and of spectral problems in general. 


{\bf Quasi-exactly solvable models} ---
About 35 years ago, a new class of spectral problems, so-called the  quasi-exactly solvable
(QES) models,  have been found for the
Schr\"odinger equation.  These are systems intermediate
to the ES systems and the non solvable ones in that one can only
determine algebraically a
part of the spectrum of the system.
\cite{TU,Tur1,Tur2,Tur3,Ush1,Ush2,Shi,O1,O2,ST}. The first and simplest QES model
discovered is the sextic oscillator \cite{TU}. The discovery of
this class of spectral problems has greatly enlarged the number of
physical systems which we can study analytically.  
Physical examples of QES systems have been found, for examples, in the motion of particles in various external fields \cite{SG,Tau,WZ,VP,HK,CH1,HR2,BP} and in nuclear physics \cite{Bohr1,Bohr2}.


{\bf Quasinormal modes} ---
The study of black hole has motivated a different kind of spectral problem, which  concerns  Hermitian Hamiltonians with complex eigenvalues.  
These eigenfunctions are called quasinormal  modes (QNMs) \cite{QNM1,QNM2,QNM3,QNM4}.  While QNMs in typical black holes cannot be solved exactly, ES and QES models with QNMs have been considered as simple examples for a better understanding of QNMs in actual black holes. 


{\bf Non-Hermitian quantum systems} ---
 About 25 years ago, it was realized that with properly defined boundary
conditions the spectrum of the non-Hermitian  $\cal{PT}$-symmetric Hamiltonian
$H=p^2+x^2(ix)^\nu$ ($\nu\geq 0$) is real and positive \cite{B1}.
Later, it was found
that a QES $\cal{PT}$-symmetric potential
can be quartic in its variable instead of sextic as in the Hermitian case
\cite{B2}.

The discovery of these non-Hermitian systems with real eigenvalues immediately sparked great interest in searching for
new systems with such properties (for a review, see e.g.,
Ref.~\cite{B3}).  While some fundamental issues remain to be addressed, non-Hermitian physics 
has already found interesting applications in many areas, such as optics, condensed matters, etc. \cite{Kwa,Oku1,Lon,MBL,Yan,Oku2}.


{\bf Rationally extended potentials} ---
Undoubtedly  one of the most
exciting developments in mathematical physics in the last 15  years has been the
discovery of  the  exceptional orthogonal polynomials \cite{GKM1,GKM2,GKM3}, and the quantal systems
related to them \cite{Que1,Que2,Que3,OS1,OS2,Roy1,Roy2,Ho7,HOS,OS3,KMR,BY,Que4}.  These new polynomials  all start with non-zero degrees  (i.e., the lowest degree member is not a constant), and yet they still form complete sets with respect to some positive-definite measure.

The quantal systems with these new exceptional orthogonal polynomials as part of their eigenfunctions are certain rational extensions  of the traditional ones related to the classical orthogonal polynomials.  
Rational extensions of solvable systems not related to the exceptional polynomials are also realized \cite{Gran1,Gran2}. In these systems, the polynomial part of the wave functions begin with degree zero.

Systematic ways to treat the above-mentioned spectral problems have been developed. 
A general and elegant way to generate the one-dimensional ES models and their rational
extensions is by the Darboux-Crum transformation
(more commonly known as the supersymmetric method in physics) together with the related idea of shape
invariance \cite{SUSY}. 
For QES systems, the established methods are the Lie-algebraic \cite{Tur1,Tur2} and the Bethe ansatz approach \cite{Ush1,Ush2}.

Recently, we have proposed an approach, which we call the prepotential approach, that can treat the ES and the QES problems in a unified way, without the need of the Darboux-Crum transformation, shape invariance, and Lie-algebra \cite{Ho1,Ho2,Ho3,Ho4}. 
In this approach  the form of
the potential of the system and the required change of variables need not be assumed from
some known systems, as in other approaches. The
potential, the required change of variables, as well as the 
eigenfunctions and eigenvalues are determined by two polynomials and a set of Bethe ansatz equations (BAEs). 
With some modifications, we have extended the approach to the rationally extended systems [5,6].

In this review the main ideas of the prepotential approach are  presented.  It is also shown how this approach can easily produce some of the well-known ES/QES Hermitian and non-Hermitian systems.  Extension to the rationally extended systems 
is briefly mentioned.

Since the idea of quasi-exact solvability is relatively new,  I shall explain
 briefly the difference between QES and ES models before discussing the prepotential
approach.


\section{Exact and quasi-exact solvabilities}

Quasi-exact solvability is most easily explained as
follows \cite{Ush2}.  The Hamiltonian $H$ of a quantal system can
be represented as a Hermitian matrix.
 The solution of the
spectral problem then reduces to the diagonalization of the matrix
$H$. This can always be done if $H$ is finite-dimensional. The system is therefore ES.
On the contrary,  there is no general algebraic rules that would allow one
to diagonalize an infinite-dimensional $H$.

If an infinite dimensional matrix $H$ is such that it can indeed  be
reduced to the diagonal form with the aid of an algebraic process,
then  the quantal system is exactly solvable.  Otherwise, the system is not exactly solvable.
 
But suppose a Hamiltonian $H$ is reducible to a block form
\begin{eqnarray}
H=\left[
\begin{array}{cccccccc}
H_{00} & H_{01} &\cdots
&H_{0N} & 0 & 0 &\cdots & 0   \\
H_{10} &H_{11} &\cdots
 &H_{1N}& 0 & 0 &\dots  & 0  \\
\cdots &\cdots &\cdots&\cdots&\cdots &\cdots &\cdots
&\cdots   \\
H_{N0}& H_{N1}&\cdots
&H_{NN} & 0 & 0 &\dots  & 0  \\
0 & 0 & \cdots &0 &*\,\,  &*\,\,  &\cdots &\cdots\\
0 & 0 & \cdots &0 &*\,\,  &*\,\,  &\cdots  &\cdots\\
\cdots &\cdots   &\cdots  &\cdots &*\,\,  &*\,\,  &\cdots &\cdots\\
0 & 0 & \cdots &0 &*\,\,  &*\,\,  &\cdots  &\cdots
\end{array}
\right],
\end{eqnarray}
where the block in the upper left corner is an $(N+1)\times (N+1)$
Hermitian matrix, and the block  with asterisks is an infinite
matrix with non-vanishing elements.  The upper left block can then be
diagonalized without touching the lower right infinite part.  Thus  a part of the spectrum of $H$ with $N+1$
eigenvalues and eigenfunctions can be determined.  This system is called QES.

The QES problem can also be understood from the viewpoint of wave mechanics as follows.
Consider a Schr\"odinger equation $H\phi=E\phi$ with
Hamiltonian $H=-d^2/dx^2 + V(x)$ and wave function $\phi (x)$.
Here $x$ belongs either to the interval $(-\infty,\infty)$,
$[0,\infty)$, or some finite interval.   The eigenstates of the most commonly known quantal systems have the form
$\phi (x)=  e^{-W_0(x)} p(z)$, where the factor $\exp(-W_0(x))$ is responsible for
the asymptotic behaviors of the wave function so as to ensure 
normalizability, and $p(z)$ is a polynomial in a new variable $z(x)$.
 The function $p(z)$ satisfies a
Schr\"odinger equation with a gauge transformed Hamiltonian
$H_W=e^{W_0} H e^{-W_0}$. 

For Hamiltonian $H$ with infinitely many eigenfunctions,  if there exists a largest integer $N$ such that the polynomial space 
${\cal V}_N=\langle 1,z,z^2,\ldots, z^N\rangle$ spanned by $\{z^k, k=0,\ldots,n\}$
is invariant under $H_W$, i.e.,
\be
H_W \cal{V}_N \subseteq \cal{V}_N,
\label{V_N}
\ee
then $H_W$ is diagonalizable in this polynomial space, and $p(z)\in \cal{V}_N$.   The system described by $H$  is then a QES system with $N+1$ eigenstates.  The polynomial space ${\cal V}_N$ is the exactly solvable sector of the system.  If (\ref{V_N}) is true for any integer $N$, then $H$ describes an ES system.  Eq.\,(\ref{V_N}) gives a criterion to check if a Hamiltonian is ES or QES.  And this we use in the Appendix to classify a Hamiltonian in the prepotential approach. 

Supersymmetric method, or Darboux-Crum transformation, is one of the most elegant methods  that treats most of the known one-dimensional ES quantum systems in a unified way \cite{SUSY}.  The success of this method turns out to rely on the amazing fact that these one-dimensional ES systems happened to possess a nice property, namely, shape invariance.  In the supersymmetric method, shape invariance is assumed as a sufficient condition to obtain the ES systems.

The main approaches to one-dimensional QES systems are the Lie-algebraic and the Bethe ansatz approach. 
The Lie-algebraic approach determines the forms of the gauge transformed Hamiltonian $H_W$  that are expressible in terms of the generators of the $sl(2)$ Lie-algebra \cite{Tur2,O1}. 
In the Bethe ansatz approach one assumes the form of the wave function containing some parameters, and then
fit these parameters to make the ansatz compatible with the chosen
potential under consideration \cite{Ush2}.
The Lie-algebraic approach to QES models excels in revealing the
underlying symmetry of a QES system explicitly.  However,
solutions of QES states are more directly found in the analytic
approach based on the BAEs .  

As mentioned in the Introduction, the prepotential approach we proposed can treat both the ES and the QES problems in a unified way.  The form of the potential,  shape invariance and the underlying symmetry of the system are not assumed, and the Darboux-Crum transformation is not used. 
 
 In the next four sections I review the main ideas of the prepotential approach
and show how some paradigmatic examples can be derived very easily. 
 Extension to the rational type is briefly discussed in Sect.\,7.

\section{Prepotential approach}

This section explains the rationale behind the prepotential approach first presented in \cite{Ho1} and later extended to QNMs in \cite{Ho4}. 

Let us begin with a one-dimensional Schr\"odinger equation with a Hamiltonian $H_0=-d^2/dx^2 +V_0(x)$ (we adopt the unit system in which
$\hbar$ and the mass $m$ of the particle are such that
$\hbar=2m=1$).   Suppose  $\phi_0(x)$ is the ground state with zero energy, i.e. $H_0\phi_0=0$.  As $\phi_0(x)$ is nodeless it can be expressed in terms of a regular function $W_0(x)$ of $x$  as $\phi_0(x)\equiv e^{-W_0(x)}$.
The Schr\"odinger equation then implies $V_0={W_0^\prime}^2 - W_0^{\prime\prime}$, where  the prime denotes derivative with respect to the variable $x$.  The potential $V_0$ is completely determined by $W_0$, hence
we shall call $W_0(x)$ the zero-th order prepotential.

In other approaches that involve the supersymmetric method, the prepotential $W_0(x)$ have been employed usually in the form of its derivative, i.e. $W^\prime_0(x)$, which is called the superpotential.  But in these methods, the form of the prepotentials (superpotential) is pre-assumed for each of the known ES models.  In our approach $W_0(x)$ is derived based on two polynomials, as shown below.

Now consider a wave function $\phi_N(x)=\phi_0(x) p_N(z)$ ($N\geq 0$), where
$p_N(z)=(z-z_1)(z-z_2)\cdots
(z-z_N), p_0\equiv 1$. Here $z=z(x)$ is some function of $x$ to be determined later. The function
$p_N(z)$ is a polynomial in an $(N+1)$-dimensional Hilbert
space with the basis $\langle 1,z,z^2,\ldots,z^N \rangle$, and $z_k$'s are its roots.  The wave function $\phi_N$ can be written as  $\phi_N=\exp(- W_N(x,\{z_k\}))$  in terms of the $N$-th order prepotential,
\be
W_N(x,\{z_k\}) = W_0(x) - \sum_{k=1}^N \ln |z(x)-z_k|. \label{W}
\ee
It satisfies a Schr\"odinger equation $H_N\phi_N=0$ with the Hamiltonian
\bea
H_N =-\frac{d^2}{dx^2} + V_N,\label{H_N}\\
V_N=
W_0^{\prime 2} - W_0^{\prime\prime}-2\left(W_0^\prime z^\prime
-\frac{z^{\prime\prime}}{2}\right)\sum_{k=1}^N \frac{1}{z-z_k} +
\sum_{k,l\atop k\neq l} \frac{z^{\prime 2}}{(z-z_k)(z-z_l)}.\n
\eea
 The potential $V_N$ is determined by $W_0(x), z^{\prime 2}$, and
the set of roots $z_k$'s ($z^{\prime\prime}$ is related to $z^{\prime 2}$ by $2z^{\prime\prime}=dz^{\prime 2}/dz$).

It will be seen that the choice of
$z^{\prime 2}$ and $W_0^\prime z^\prime$ determine the nature of
solvability of the quantal system. 
Here we
consider only those cases where $W_0^\prime z^\prime=P_m
(z)$ and $z^{\prime 2}=Q_n(z)$ are
polynomials in $z$ of degree $m$ and $n$, respectively.  The variables $x(z)$ (assumed invertible for practical purposes) and the prepotential $W_0(x)$ are given by
\bea
x(z) &=&\pm \int^z \frac{dz}{\sqrt{Q_n(z)}},\n\\
W_0(x) &=& \int^x dx\,\left(\frac{P_m(z)}{\sqrt{Q_n(z)}}\right)_{z(x)}
=\left(\int^z\,dz\,\frac{P_m(z)}{Q_n(z)}\right)_{z(x)}. 
\label{W0}
\eea
The second equality in (\ref{W0}) is obtained using $W_0^\prime = z^{\prime}dW_0/dz$.
The two relations in (\ref{W0}) define the new coordinate $z(x)$
and the corresponding prepotential $W_0(x)$. Thus, $P_m(z)$ and
$Q_n(z)$ determine the Hamiltonian $H_0$.  Of course, the choice of
$P_m(z)$ and $Q_n(z)$ should  ensure normalizability of $\phi_0=\exp(-W_0)$.

Analysis in the Appendix shows that, depending on the degrees of the polynomials $P_m(z)$ and $Q_n(z)$,
we have the following situations \footnote{We take this opportunity to correct the values of $m, n$ for the case (ii) and (iii) in \cite{Ho1,Ho4}.}:

\begin{enumerate}
\item[(i)] if $m\leq 1, n\leq 2$, then in $V_N(x)$
the parameter $N$ and the roots $z_k$'s will only appear as an
additive constant and not in any term involving powers of $z$.
There is an invariant subspace of $H$ for any $N$. 
Such system is then exactly solvable. The additive constant gives the eigenvalue.

\item[(ii)] if $m=2, n\leq 2$, then $N$ will appear in
the first power term in $z$, but $z_k$'s only in an additive term in $V_N$.
For each $N\geq 0$, there is a $(N+1)$-dimensional invariant subspace.
Hence, the system with $N$-dependent potential is QES with $N+1$ solvable states.  

\item[(iii)] if $m {\rm\ or\ } n>2$, then not only $N$ but also
$z_k$'s will appear in the coefficients of powers of $z$ in the potential. 
There is no invariant subspace in this case.   
Each set of the roots  defines a QES potential with only one eigenfunction 
$e^{-W_0} p_N(z)$.

\end{enumerate}

Case (iii) is not of much interest here. Henceforth we shall consider only cases with both
$m,~n\leq 2$, i.e., Case (i) and (ii). Let $P_2(z)=A_2 z^2 + A_1 z + A_0$ and $Q_2(z)=q_2
z^2 + q_1 z + q_0$.  The solvability of the system is
determined solely by $A_2$: ES if $A_2=0$,  QES otherwise. 
Coordinates $z(x)$ with $z^{\prime 2}=Q_2(z)$ quadratic in $z$
are called  ``sinusoidal coordinates", which include
quadratic polynomials, trigonometric, hyperbolic, and exponential
types.  This choice of $z^{\prime 2}$ covers most of
the known ES shape-invariant potentials \cite{SUSY}
and the $sl(2)$-based QES systems \cite{Tur2}.

We can recast $V_N$ into the form \cite{Ho1}
\bea
 V_N &=&{W_0^\prime}^2 - W_0^{\prime\prime}+ q_2 N^2 -2 A_1N
 - 2 A_2 N z - 2A_2\sum_{k=1}^N z_k \\ \n
&& -2\sum_{k=1}^N
\frac{1}{z-z_k}\left\{P_2(z_k)-\frac{q_2}{2}z_k -
\frac{q_1}{4} - \sum_{l\neq k} \frac{Q_2(z_k)}{z_k-z_l}\right\}.
\label{V-1}
\eea
 $V_N$ is free of simple poles if all the residues at $z_k$'s (i.e., terms in the bracket) vanish.  This gives the BAEs
satisfied by the roots $z_k$'s:
\begin{equation}
A_2 z_k^2 + \left(A_1 - \frac{q_2}{2}\right) z_k +
A_0-\frac{q_1}{4}-\sum_{l\neq k}\frac{q_2 z_k^2 + q_1 z_k
+q_0}{z_k-z_l}=0.\label{BAE}
\end{equation}
Finally, after some algebra, 
we arrive at the potential \cite{Ho4}
\bea
V_N (x) 
&=& \frac{P_2^2 - Q_2 \frac{dP_2}{dz} + \frac12 P_2
\frac{dQ_2}{dz}}{Q_2}-\left(2 A_1N - q_2 N^2
 +2 A_2 N z + 2A_2\sum_{k=1}^N z_k\right)\nonumber\\
&=&\left[\frac{(A_2 z^2 + A_1 z + A_0)^2}{q_2 z^2 + q_1 z + q_0} -2(N+1)A_2z \right. \n \\
&&\left.  +\frac12\left(A_2 z^2 + A_1 z +
A_0\right)\frac{2q_2 z + q_1}{q_2 z^2 + q_1 z + q_0}\right]_{z(x)}\n \\
&& -\left[(2N +1)A_1- q_2 N^2
  + 2A_2\sum_{k=1}^N z_k\right],
 \label{V-2}
\eea
and the wave function
\begin{eqnarray}
\phi_N \sim e^{-\int^{z(x)} dz
\frac{P_2(z)}{Q_2(z)}}\,p_N(z).
\label{wf-N}
\end{eqnarray}
Eqs.\,(\ref{BAE}), (\ref{V-2}) and (\ref{wf-N}) define the most general ES and QES systems based on
sinusoidal coordinates.

In \cite{Ho1} systems defined on the whole line, the half-line, and finite interval are treated separately.  It was soon realized that this was unnecessary.  The domain of the original variable $x$ is indeed determined by the singularities of $V_N(x)$ not related to the roots of $p_N(z)$, such as $x=0, \pm\infty$.  

The sinusoidal coordinates come in three inequivalent canonical forms \cite{Ho3,O2}:
\be
  z^{\prime 2}=\left\{ \begin{array}{ll}
              \gamma \neq 0 &\mbox{(harmonic oscillator)~;}\\
               \beta z (\beta>0) & \mbox{(radial oscillator)~;}\\
               \alpha (z^2  + \delta) & \mbox{(Scarf I, II; Morse; P\"oschl-Teller)~;}
               \end{array}
               \right.
\ee
Examples of models with these forms of coordinates have been considered in \cite{Ho1,Ho4}. For clarity of presentation, in what follows  we will discuss only systems  of the harmonic and the radial oscillators types.

\section{Hermitian potentials with real/complex spectra I ($n=0 : z^{\prime 2}=\gamma>0$)}

The transformation corresponding to the choice $z^{\prime 2}=\gamma>0$  is
$z(x)=\sqrt{\gamma}x+\rm{constant}$. Without loss of generality,  we shall take
$z(x)=x$. The potential $V_N$ and the ground state $\phi_0\sim\exp(-W_0)$ of this system are \cite{Ho4}
\begin{eqnarray}
V_N &=&A_2^2  x^4 + 2 A_2A_1 x^3 +\left(A_1^2 +
2A_2A_0\right)x^2 + 2\left[A_1A_0
-A_2(N+1)\right] x \n\\
&&-\left[2A_1(N+\frac12)-A_0^2+ 2A_2\sum_{k=1}^N
z_k\right]
\label{V1}
\end{eqnarray}
and 
\begin{eqnarray}
\phi_0(x)\sim e^{-\frac13 A_2 x^3 -\frac12 A_1 x^2 - A_0 x}.
\label{wf0}
\end{eqnarray}
The only singularities of $V_N(x)$ are $x=\pm\infty$, so the domain of $x$ is ($-\infty, \infty$).

\subsection{$m=1$: simple harmonic oscillator}

As an example, let us take $A_2=A_0=0$ and $A_1=b$, i.e.,  $P_1(z)=bz$. 
 Eq.~(\ref{wf0}) gives the ground state
\begin{eqnarray}
\phi_0(x)\sim e^{ -\frac12 bx^2 }.
\end{eqnarray}
 One must assume $b>0$ in order for $\phi_0(x)$ to be
square-integrable. The BAEs (\ref{BAE}) are
\begin{eqnarray}
bx_k-\sum_{j\neq k} \frac{1}{x_k-x_j}=0, ~~k=1,\ldots,N,
\label{BAE-Osc}
\end{eqnarray}
and the potential is $V_N=b^2 x^2 -(2N+1)b$. 
Note that $N$ appears only in the additive constant term, and the roots $z_k$'s do not appear at all (as $A_2=0$). 
 If we define the potential of the system only by the first term in $V_N$, then we have the Schr\"odinger equation:
\begin{eqnarray}
\left(-\frac{d^2}{dx^2} + b^2 x^2 \right)e^{-W_N}=b(2N+1)e^{-W_N}.
\end{eqnarray}
  This system is just the well-known simple
harmonic oscillator.

Note that the rescaling $\sqrt{b}x_k\to x_k$ will set $b=1$ in
Eq.~(\ref{BAE-Osc}).  This is the BAEs that determine the zeros of the Hermite polynomials
$H_N(x)$ as found by Stieltjes. Hence
the well known wave functions for the harmonic
oscillator, namely, $\phi_N=\exp(-W_N)\sim
\exp(-bx^2/2)H_N(\sqrt{b}x)$, are reproduced.

\subsection{$m=1$ : QNM}

The choice $A_0=0$ and $A_1=-i\frac{c}{2}$ gives the simplest exactly solvable QNM model with potential \cite{Ho4}
\begin{equation}
V_N=-\frac{1}{4}c^2x^2 + ic \left(N+\frac12\right).
\end{equation}
The imaginary part of the eigenvalue is
proportional to $N+1/2$, which is characteristic of most black hole
QNMs.

\subsection{$m=2$: QES model}
We want $V_N$ to be real. If $A_2\neq 0$, then the term $A_2(N+1)$
in the fourth term of (\ref{V1})  implies $A_2$ be real. However, this implies
the wave function $\phi_N$, whose asymptotic behavior is governed
by $\phi_0$ in (\ref{wf0}), is not normalizable on the whole line. Hence, in this case  there is
no QES models, with or without QNMs.

\section{Hermitian potentials with real/complex spectra II ($n=1: z^{\prime 2}=\beta z,~\beta>0$)}

In this case we can take  $z(x)=\frac{\beta}{4}x^2$.  The potential and the ground state are \cite{Ho4}
\begin{eqnarray}
V_N &=& \frac{1}{64}A_2^2\beta^2 x^6 + \frac{1}{8}A_2A_1\beta x^4
+\frac{1}{4}\left[A_1^2+
2A_2A_0-{A_2\beta}\left(2N+\frac32\right)
\right]x^2\nonumber\\
&&
+\frac{4A_0}{\beta}\left(\frac{A_0}{\beta}+\frac{1}{2}\right)\frac{1}{x^2}
-\left[A_1\left(2N+\frac12-\frac{2A_0}{\beta}\right)+2A_2\sum_{k=1}^N
z_k\right] \label{V-3d-osc}
\end{eqnarray}
and
\begin{equation}
\phi_0(x)\sim
x^{-\frac{2A_0}{\beta}}e^{-\frac{1}{4}A_1x^2-\frac{1}{32}A_2\beta x^4}.
\end{equation}
The singularities of $V_N(x)$ are $x=0, \infty$, so the domain of $x$ is ($0, \infty$).

\subsection{$m=1$: radial oscillator}

For $A_2=0$ with real values of $A_1$ and $A_0$, $V_N$ is the potential of the radial oscillator.
Particularly, letting $A_1=2a>0, A_0=-2\ell$ and $\beta=4$ with real parameters $a$ and $\ell$, 
(\ref{V-3d-osc}) becomes
\begin{equation}
V_N=a^2x^2 +\frac{\ell(\ell - 1)}{x^2}  - a \left(4N+2\ell
+1\right).
\end{equation}
The additive constant gives the eigenvalue for each $N$.
The ground state is
\be
\phi_0(x)\sim x^\ell e^{-\frac12 a x^2}
\ee
The corresponding set of BAEs  is
\begin{eqnarray}
2az_k +2\ell +1 + 4 \sum_{j\neq k} \frac{z_k}{z_k-z_j}=0, ~~k=1,\ldots,N,
\label{BAE-3dOsc}
\end{eqnarray}

\subsection{$m=1$: QNM}

The choice  $\beta=4, A_0=-2\ell$, and $A_1=-2ia$ defines
a simple model with QNMs related to the radial oscillator \cite{Ho4}:
\begin{equation}
V_N=-a^2 x^2 +\frac{\ell(\ell - 1)}{x^2}  +i a \left(4N+2\ell
+1\right).
\end{equation} 
The BAEs are obtained by simply making the change $a\to -ia$ in (\ref{BAE-3dOsc}).

\subsection{$m=2$: QES sextic oscillator}

It is obvious from (\ref{V-3d-osc}) 
that if $A_2\neq 0$, then all $A_i$'s have to be
real. Hence there is no QES model with QNMs in this case. For
$A_2>0$ and $A_0<0$, the wave functions $\phi_N$ are normalizable
on the positive half-line. The system is then a QES model with
real energies.  Together with the discussion in the last section, one
concludes that one-dimensional QES models start with degree six,
i.e., the sextic oscillator.   In fact, the first three terms in the potential 
(\ref{V-3d-osc}) with $A_2=2a>0,~A_1=2b,~A_0=0$ and $\beta=4$,
namely,
\begin{eqnarray}
V_N=a^2 x^6 +2ab x^4+ \left[b^2-(4N+3)a\right]x^2,
\label{sextic}
\end{eqnarray}
with eigenvalue $E= (4N+1)b+4a{\sum_k z_k}$, is the very first QES model discussed in the literature \cite{TU}.
Note the appearance of $N$ in the $x^2$-term.
The BAEs (\ref{BAE}) are:
\begin{eqnarray}
2az_k^2 +2bz_k -1 - 4\sum_{l\neq k}\frac{z_k}{z_k-z_l} =0,~~~
k=1,\ldots,N,
\label{BAE-Q1}
\end{eqnarray}

It is instructive to see how some lowest states of this model are determined in our approach.

For  $N=0$, the potential is just $V_0$ with only one solvable state $\phi_0=\exp(-a x^4/4-bx^2/2)$ corresponding to energy $b$.

The solutions of the BAEs for $N=1$ are $z_{1\pm}=(-b\pm \sqrt{b^2+2a})/2a$ and the energies are $E_{1\pm}=3b\pm 2\sqrt{b^2+2a}$.
The two corresponding orthogonal eigenfunctions are $\phi_{1\pm}(x)=\phi_0(x)(z-z_{1\pm})$.  These results are consistent with the Lie-algebraic solutions \cite{Ush2}.

The BAEs (\ref{BAE-Q1}) for the case $N=2$ can be conveniently handled as follows. By adding and subtracting the two BAEs for $k=1 $ and $k=2$,  we can write the BAEs as
\be
a(p^2+ q^2) + 2b p -6=0, ~~(ap+b)q^2-2p=0,
\ee
where $p\equiv z_1+z_2$ and $q\equiv z_1-z_2$.
These equations can be further reduced to
\bea
a^2 p^3  + 3 a b p^2 + \left( 2b^2-4a \right) p - 6b = 0,~~~q^2 = \frac{2p}{ap+b}.
\label{pq}
\eea
Solving the first equation in (\ref{pq}) for $p$  then gives the eigenvalues and eigenfunctions
\bea
E=9b+ 4ap, ~~ \phi_2(x)=\exp\left(-\frac14 ax^4 - \frac12 bx^2\right) \left(z^2 - p z  + \frac{p^2-q^2}{4}\right).
\eea

While analytic solutions of the BAEs for  general  $a$ and $b$ can be obtained, they are rather complicated.  Below we shall only present the eigenvalues and the orthogonal eigenfunctions  for the case with $a>0, b=0$ and $a=b=1$.
\begin{enumerate}
\item[a)]~~ $a>0,~b=0: p=0, \pm 2/\sqrt{a} $
\bea
E=0, &&~~~ \phi_{20}(x)=e^{-\frac14 a x^4} \left(x^4-\frac{3}{2a}\right),\\
E= \pm 8\sqrt{a}, &&~~~ \phi_{2\pm}(x)=e^{-\frac14 a x^4} \left(x^4 \mp \frac{2}{\sqrt{a}} x^2+\frac{1}{2a}\right).\n
\eea
\item[b)]~~ $a=b=1: p=-3, \pm\sqrt{2}$
\bea
E=-3, &&~~ \phi_{20}(x)=e^{-\frac14  x^4-\frac12 x^2} \left(x^4+3x^2+\frac{3}{2}\right),\label{phi2}\\
E= 9 \pm 4 \sqrt{2}, &&~~\phi_{2\pm}(x)=e^{-\frac14  x^4-\frac12 x^2} \left(x^4 \mp \sqrt{2} x^2 \pm \frac{\sqrt{2}\mp 1}{2}\right).\n
\eea
\end{enumerate} 

The states with $E=-8\sqrt{a}$ and $E=-3$ above are the ground states, as they have no nodes in the physical domain of $x$.
For case (a) our results are consistent with those obtained by the algebraic method for $a=1, b=0$ in \cite{Shi}. 
 
We note here that the case $N=2, a=1$ and $b=-1$ can be obtained from the results in Case (b) above by simply making the change $p\to -p,~q^2\to q^2$.

In Fig.\,1 we show some plots of the sextic potential and the QES states for $a=b=1$ and $N=1,2$. 

For higher values of $N$, analytic solution of the BAEs becomes very complicated, and numerical method is needed.


\begin{figure}[ht] \centering
\includegraphics*[width=7cm,height=6cm]{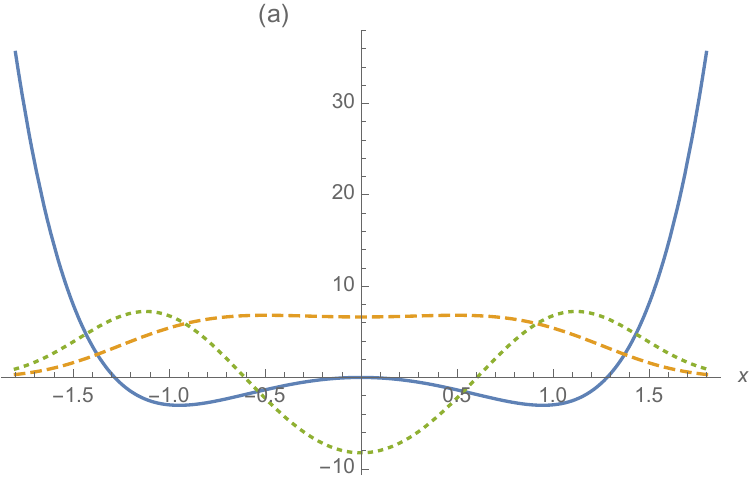}\hspace{1cm}
\includegraphics*[width=7cm,height=6cm]{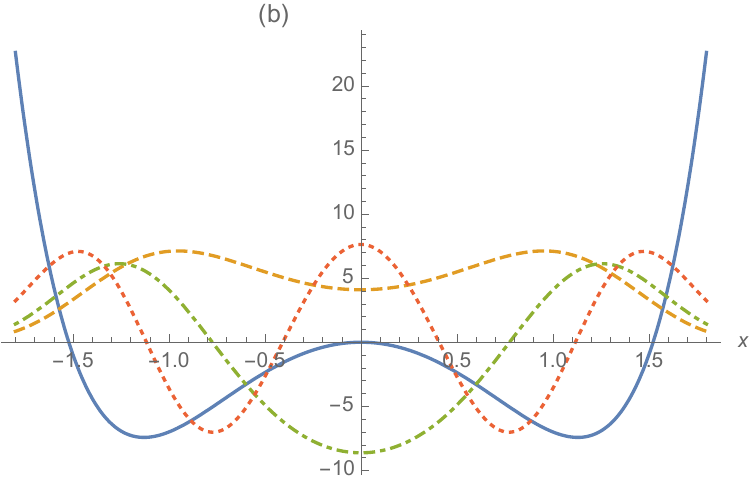}
\caption{Plots of  the sextic oscillator potential $V_N$ (\ref{sextic})  and its $N+1$ normalized QES states (magnified  $10$ times) for $a=b=1$  :  (a) For $V_1(x)$ (solid), $\phi_{1-}(x)$ (dashed), and $\phi_{1+}(x)$ (dotted); (b) For $V_2(x)$ (solid), $\phi_{20}(x)$ (dashed), $\phi_{2-}(x)$ (dotdashed), and $\phi_{2+}(x)$ (dotted) in (\ref{phi2}).
}
\label{Fig1}
\end{figure}

\section{Non-Hermitian potentials with real spectra}

In the previous section we have seen that the lowest-degree
one-dimensional QES polynomial potential is sextic.  But as mentioned in the Introduction,  
it was found that, by 
allowing non-Hermitian $\cal{PT}$-symmetric Hamiltonians,
a QES polynomial potential can be quartic in its variable
\cite{B2}.  Now let us demonstrate how simply the
Hamiltonian discovered in Ref.~\cite{B2} is derived in our
approach if we allow $P_m(z)$, or equivalently $W_0(x)$, to be
complex.


Recall we mentioned  in Sect.\,4.3 that
there is no  QES Hamiltonian with
quartic polynomial potential if all the $A_k$'s in $P_2(x)$ are real.  

The situation is different if we allow $P_2(x)$ to be complex.  Let
$A_2=i\alpha,~A_1=\beta$ and $A_0=i\gamma$, where $\alpha,~\beta>0$ and
$\gamma$ are real.   Now $\exp(-W_0)$ in (\ref{wf0}) is
square-integrable on the whole line for $\beta>0$. The potential and BAEs are obtained from
Eqs.~(\ref{V1}) and (\ref{BAE}) (with $q_2=q_1=0, q_0=1$) to be
\begin{eqnarray}
V_N(x)&=&-\alpha^2 x^4 + 2i\alpha \beta x^3 + \left(\beta^2
-2\alpha \gamma\right)x^2
-2i\left[(N+1)\alpha-\beta\gamma\right]x\nonumber\\
&& - \left[\beta (2N+1) + \gamma^2+2i\alpha\sum_k x_k\right]
\label{complex1}
\end{eqnarray}
and
\begin{eqnarray}
i\alpha x_k^2+\beta x_k + i\gamma - \sum_{l\neq k}
\frac{1}{x_k-x_l}=0,~~k=1,2,\ldots,N.
 \label{BAE-complex1}
\end{eqnarray}
The potential given by the first four terms in Eq.~(\ref{complex1}) is precisely the
$\cal{PT}$-symmetric QES quartic potential obtained in
Ref.~\cite{B2} (in the notation of Ref.~\cite{B2}, the
potential has parameters $\alpha=1$, $\beta=a$, $\gamma=b$, and
$N+1=J$).   It is nice to see that in our
prepotential approach this potential can be so simply and directly
derived, without any knowledge of its underlying symmetries.

The properties of this system have been described in
Ref.~\cite{B2}.  So here we just consider the two simplest cases with $N=0$ and
$N=1$.

For $N=0$, the potential is just $V_0$ with only one solvable
state corresponding to energy $E=0$.  For $N=1$, there are two
solvable states.  As with Ref.\,\cite{B2}, we define the QES potential by the first four terms in Eq.~(\ref{complex1}):
\begin{equation}
V_1=-\alpha^2 x^4 + 2i\alpha\beta x^3 + \left(\beta^2
-2\alpha\gamma\right)x^2 -2i(2\alpha-\beta\gamma)x.
\end{equation}
The eigenvalues are $\gamma^2 +3\beta+2i\alpha x_1$.  The roots $x_1$
satisfy the BAEs (\ref{BAE-complex1}): $i\alpha x_1^2+\beta x_1 +
i\gamma=0$. The solutions are $x_1=i(\beta\pm
\sqrt{\beta^2+4\alpha\gamma})/2\alpha$.  It is seen that for
$\beta^2+4\alpha\gamma >0$ the non-Hermitian potential $V_1$ has
two normalizable states with real energies $\gamma^2+ 2\beta \pm
\sqrt{\beta^2+4\alpha\gamma}$.  But if $\beta^2+4\alpha\gamma <0$,
the energies become complex, and the $\cal PT$ symmetry is said to be broken. This is consistent with the numerical
results in Ref.~\cite{B2}.  Analysis for higher values of $N$ becomes more complicated and numerics are needed \cite{B2}.

We mention here that a new non-$\cal PT$ case is obtained if we
assume the parameter $b$ in the sextic oscillator  in Sect.\,5.3. to be purely imaginary: $b=
i\beta$, where $\beta$ and $a>0$ are real.  We will not discuss this case in detail owing to space limitation.

\section{Rationally extended potentials}

Complexity involved in the rationally extended models no longer allows us to derive them from  such simple assumptions as the forms of some polynomials, like $P_m(z)$ and $Q_n(z)$  in the non-rational cases.  New inputs are needed. 
The main ideas of how the prepotential approach is extended to the rationally extended models are briefly summarized below. 
We refer the reader to \cite{Ho5,Ho6} for the details of the procedure. \footnote{We note here that prepotential used in \cite{Ho5,Ho6} differ from the one here by a negative sign.  This is to facilitate comparison with the results in some earlier work on exceptional orthogonal polynomials, particularly Refs. \cite{OS1,OS2}.}

The wave functions in these models have the form
\begin{eqnarray}
\phi(x) =\frac{e^{-W_0(x)}}{\xi(z)}\,p(z), \label{phi}
\end{eqnarray}
which involve a polynomial $\xi(z)$, the deforming function,  in the denominator. 
  The presence of $\xi(z)$ in the
denominators of $\phi(x)$ and the potential thus gives a rational
extension, or deformation, of the traditional system.  
For $\xi(z)=1$ the model reduces to the corresponding traditional ES system.
To generate rational ES systems, it is assumed that the prepotential $W_0$ be a regular function of $x$,  the zeros of the function $\xi(z)$  lie outside the
physical domain of $z(x)$, and the function $p(z)$ does not appear in the potential.

In Ref. \cite{Ho5, Ho6} we have given a procedure to determine $W_0(x),~\xi(z)$ and $ p(z)$.
The prepotential is assumed to have the form
$
W(x,z) = W_0(x)+\ln \xi(z) - \ln p(z)
$
so that $\phi (x)= \exp(-W(x))$.
The new variable $z(x)$ is chosen to be
one of the sinusoidal coordinates.
The function $p(z)$ is assumed to be  a linear combination of
$\xi(z)$ and $d\xi(z)/dz$:
\begin{eqnarray}
p(z)= G(z) \xi(z)+ F(z) \frac{d}{dz}\xi(z).
 \end{eqnarray}
The functions $\xi(z), F(z)$ and $G(z)$ are determined by matching the equations they satisfied with the hypergeometric equations. 
It turns out $\xi (z)$ depends on an integral index $\ell> 0$, i.e., $\xi_\ell (z)$. The system is called single-indexed rational model.

In Eq.~(\ref{phi}), $p(z)={\rm\  constant}$ is admissible
if the factor $\phi_0(x)=\exp(-W_0(x))/\xi(z)$  is
normalizable, and $\phi_0(x)$ is the ground state. Otherwise, $\phi_0(x)$ cannot be the ground state.
In this case, the ground
state, like all the excited states, must involve non-trivial
$p(z)\neq {\rm\ constant}$.  This give rise to the 
exceptional orthogonal polynomials \cite{Ho5}.  

For illustration purpose, here we only list the main data of the deformed radial oscillator,  which is the first quantum system that involves the new exceptional  orthogonal polynomials.


\begin{enumerate}

\item[-]
Deforming function:  $ \xi_\ell(z)=L_\ell^{(\alpha)}(z), ~~L_\ell^{(\alpha)}(\cdot)$: Laguerre polynomials; \\
$z(x)=x^2, ~\alpha<-\ell$ (so $\xi_\ell$ has no zeros in physical domain $[0,\infty)$).

\item[-]
$0^{th}$-order  prepotential: $ W_0(x)=-\frac{x^2}{2} + \left(\alpha+\frac12\right)\ln x$, ~~~$\phi_0$:  non-nomalizable.

\item[-]
Potential:   
$ V(x)=x^2 +
\frac{\left(\alpha+\frac12\right)\left(\alpha+\frac{3}{2}\right)}{x^2}
+8\frac{d\ln \xi_\ell}{dz}\left[z\left(\frac{d\ln \xi_\ell}{dz}
-1\right)+\alpha+\frac12\right]+ 2(2\ell-\alpha)$.

\item[-]
Eigenfunctions and energies:
\begin{eqnarray}
 \phi_{\ell,n}(x;\alpha)& \propto&
 \frac{e^{-\frac{x^2}{2}}x^{-(\alpha+\frac12)}}{\xi_\ell (z)}} { p_{\ell,n}(z), 
~~~~\mathcal{E}_{\ell,n}=4(n-\alpha-\ell) ,\n \\
 p_{\ell,n}(z)&=& (\alpha-n)L_n^{(-\alpha-1)}(z)\xi_\ell (z) +  z L_n^{(-\alpha)}(z)\frac{d}{dz}\xi_\ell (z).
\n
\end{eqnarray}

\end{enumerate}

Here $p_{\ell,n}(z)$ is of degree $\ell+n$.
For  $\ell\to  0$ , $\xi_\ell(z) \to 1$, the system reduces to the  ordinary radial oscillator!

\section{Summary}

This work presents a brief overview of the general ideas of 
the prepotential approach to the ES and QES models.  
The method is demonstrated with simple examples of 
Hermitian and non-Hermitian Hamiltonians with real energies, 
and Hermitian systems with complex energies (quasinormal modes).  Extension of the approach to the newly discovered rationally extended systems is briefly mentioned.

We have succeeded by using this approach to generate in a unified way all the known one-dimension ES quantum models related to supersymmetry, various simple systems with QNMs, some non-Hermitian systems with QES real spectra, and rational extended ES models, without exploiting any underlying symmetries of the systems \cite{Ho1,Ho2,Ho3, Ho4,Ho5,Ho6}.  Nevertheless,  what we have done so far  is still quite preliminary.  Our
approach could be further developed. 

Firstly, we could generalize the approach to
cases for which $W_0^\prime z^\prime$ and $z^{\prime
2}$ are non-polynomials in $z$.  Secondly, it is
extremely interesting to see how our formalism could be extended
to many-body systems, such as the renowned
Calogero-Sutherland-Moser systems.  The analogues of Bethe ansatz
equations for these systems are not known yet. Thirdly, although we have shown how to  generate
some non-Hermitian Hamiltonians using complexified 
$P_m(z)$, it seems necessary that one should extend the basic variable $x$ to the complex plane
for a better understanding of these Hamiltonians 
\cite{B1,B2}.  How to extend the prepotential
approach to complex variable deserves further study. 
Finally, one hopes to extend the approach to systems with multi-indexed exceptional polynomials \cite{GKM3,OS3}.

\section*{Acknowledgments}

This work is supported in part by the National Science and Technology Council (NSTC) of
the Republic of China under Grant Nos. NSTC 112-2112-M-032-007. I thank A. Moroz for helpful comments  that  have led me to re-examine the conditions on $m$ and $n$ for Case (ii) and (iii).

\bigskip


\leftline{{\bf Appendix}}

We have the Schr\"odinger equation $H_N\phi_N=0$, where $\phi_N=e^{-W_0}\,p_N$ and $H_N$ is
\bea
H_N &=&-\frac{d^2}{dx^2} + W_0^{\prime 2} - W_0^{\prime\prime}\n \\
&& -\left(2 P_m(z)
-\frac12\frac{d Q_n}{dz}\right)\sum_{k=1}^N \frac{1}{z-z_k} +
\sum_{k,l\atop k\neq l} \frac{Q_n(z)}{(z-z_k)(z-z_l)},
\eea
from Eq.\,(\ref{H_N}).
The polynomial $p_N(z)$ satisfies $H_W\,p_N=0$, where $H_W=e^{W_0}H_N e^{-W_0}$ is
\bea
H_W&=&-Q_n\frac{d^2}{dz^2}
+\left(2 P_m(z)-\frac12\frac{d Q_n}{dz}\right)\left[\frac{d}{dz}-\sum_{k=1}^N \frac{1}{z-z_k} \right]\n\\
&&+\sum_{k,l\atop k\neq l} \frac{Q_n(z)}{(z-z_k)(z-z_l)}.
\eea
The action of $H_W$ on $z^j$ is
\bea
H_W\, z^j~~\sim &&- j(j-1)Q_n z^{j-2} 
+\left(2 P_m(z)-\frac12\frac{d Q_n}{dz}\right)\left[j z^{j-1}-\sum_{k=1}^N \frac{z^j}{z-z_k} \right]\n\\
&& +\sum_{k,l\atop k\neq l} \frac{Q_n(z)}{(z-z_k)(z-z_l)}\,z^j.
\eea

If  $m\leq 1, n\leq 2$, then ${\rm deg}\,\{H_W z^j\}\leq j$.  Hence, the space $\mathcal{V}_j=\langle 1, z, z^2, \ldots, z^j\rangle$ is a $(j+1)$-dim invariant subspace of $H_N$, i.e.,
$H_W\mathcal{V}_j \subseteq \mathcal{V}_j$, for any $j$. Hence $H_N$ is exactly solvable, as $\mathcal{V}_j$ can be diagonalized for any $j$.

Next,  if  $m=2, n\leq 2$ (note: not $ {\rm max}\,\{m,n-1\}=2$ as in \cite{Ho1}). Now 
$H_W\,\mathcal{V}_j \subseteq \mathcal{V}_j$ only holds for $j\leq N$.  This is seen as follows. 
The first and the third term in (4) do not increase the degree of $z^j$ in this case, but the second term will give rise to a term $z^{j+1}$ when $m=2$, as can be seen from
\bea
&&\left(2 P_2 (z)-\frac12\frac{d Q_n}{dz}\right)\left[j z^{j-1}-\sum_{k=1}^N \frac{z^j}{z-z_k} \right]\n
\\
\sim && \left(2 P_2(z)-\frac12\frac{d Q_n}{dz}\right)\left[jz^{j-1}-N z^{j-1}+ {\rm terms\  with\  deg}<j-1 \right].
\eea
But it is fine for $j< N$ as $H_W z^j$ is still in the space $\mathcal{V}_N$, even a term in $z^{j+1}$ appears. For $j=N$, the first two terms in $z^{N-1}$ cancel, so $H_W z^N$ is still in the space $\mathcal{V}_N$ as $z^{N+1}$ doesn't appear. 
Thus in this case, for each $N$, there is a $(N+1)$-dim invariant subspace of $H_W$, and hence there exists $N+1$ exact eigenfunctions of $H_W$ (obtained by the diagonalization of $H_W$).  This means there are $N+1$ sets of roots of  the BAEs. So $H_N$ is QES. 

Finally, when $m$ or $n\geq 3$, there  is no invariant subspace of $H_N$. 

\leftline{\bf References}


\begin{thebibliography}{9}


\bibitem{Ho1}
 Ho C-L 2008 Ann. Phys. {\bf 323} 2241  

\bibitem{Ho2}
Ho C-L  2009 Ann. Phys. {\bf 324} 1095  

\bibitem{Ho3}
Ho C-L 2009 J. Math. Phys. {\bf 50} 042105 

\bibitem{Ho4}
Ho C-L 2011 Ann. Phys. {\bf 326} 1394  

\bibitem{Ho5}
Ho C-L 2011, Prog. Theor. Phys. {\bf 126} 185 

\bibitem{Ho6}
Ho C-L 2011 J. Math. Phys. {\bf 52} 122107   


\bibitem{TU}  
Turbiner A V and Ushveridze A G 1987 Phys. Lett. {\bf A126} 181

\bibitem{Tur1} 
Turbiner A V 1988 Sov. Phys. JETP {\bf 67} 230

\bibitem{Tur2} 
Turbiner A V 1988 Comm. Math. Phys. {\bf 118} 467

\bibitem{Tur3}
Turbiner A V 2016 Phys. Rep. {\bf 642} 1

\bibitem{Ush1} 
Ushveridze A G 1988  Sov. Phys.-Lebedev Inst. Rep. {\bf 2} 50; 54

\bibitem{Ush2}
Ushveridze A G  1994 {\sl Quasi-exactly Solvable Models in Quantum Mechanics} (IOP, Bristol).

\bibitem{Shi}
Shifman M A 1989 Int. J. Mod. Phys. {\bf A4} 2897

\bibitem{O1} 
Kamran N and Olver P J  1990 J. Math. Anal. Appl. {\bf 145} 342

\bibitem{O2}
Gonz\'alez-L\'opez A, Kamran N, and Olver P J 1993  Comm. Math. Phys. {\bf 153} 117

\bibitem{ST}  
Shifman M A and Turbiner A V 1989 Comm. Math. Phys. {\bf 126}  347


\bibitem{SG} 
Samanta A and Ghosh S K 1990  Phys. Rev. {\bf A42} 1178

\bibitem{Tau}
Taut M 1993 Phys. Rev.  {\bf A48} 3561

\bibitem{WZ} 
Wiegmann P B and Zabrodin A V 1994 Phys. Rev. Lett. {\bf 72} 1890

\bibitem{VP} 
Villalba V M and Pino R 1998 Phys. Lett. {\bf A238}, 49

\bibitem{HK} 
Ho C-L and Khalilov V R 2000  Phys. Rev. {\bf A61} 032104

\bibitem{CH1}
Chiang C-M and Ho C-L 2002 J. Math. Phys. {\bf 43} 43



\bibitem{HR2} 
Ho C-L and Roy P 2004 Ann. Phys. {\bf 312} 161

 \bibitem{BP}      
{Baradaran M and Panahi H 2017
Adv. High Energy Phys. {\bf 2017} 8429863}
        

\bibitem{Bohr1}         
{Sobhani H, Hassanabadi H, Bonatsos D, Pan F, Cui S , Feng Z, and  Draayer  G P 2020
Eur. Phys. J. {\bf A 56} 29}

\bibitem{Bohr2}      
{Sobhani H, Hassanabadi H, Bonatsos D, Pan F, and Draayer  G P 2020
Nucl.  Phys. {\bf  A 1002} 121956}


\bibitem{QNM1}
Chandrasekhar S 1983 {\sl The Mathematical Theory of Black Holes} (Clarendon, Oxford).

\bibitem{QNM2}
Ferrari V and Mashhoon B 1984  Phys. Rev.  {\bf D30} 295
 
\bibitem{QNM3}
Kokkotas K D and  Schmidt B G 1999  Living Rev. Rel. {\bf 2} 2

\bibitem{QNM4}
Motl L 2003 Adv. Theor. Math. Phys. {\bf 6} 1135


\bibitem{B1} 
Bender C M and  Boettcher S 1998 Phys. Rev. Lett. {\bf 80} 5234

\bibitem{B2} 
Bender C M and  Boettcher S 1998 J. Phys. {\bf A31} L273

\bibitem{B3}
Bender C M 2005 Contemp. Phys. {\bf 46} 277

 \bibitem{Kwa}   
 {Kawabata K, Higashikawa S, Gong Z, Ashida Y, and  Ueda M 2019
Nat. Commun. {\bf 10} 297}

\bibitem{Oku1}
{Okuma N, Sato M  2019 Phys. Rev. Lett. {\bf 123} 097701}

\bibitem{Lon}   
{Longhi  S 2019 Phys. Rev. Lett. {\bf 122} 237601}
 
 \bibitem{MBL}
{Marie A, Burton H G A, and  Loos P-F 2021  J. Phys.: Condens. Matter {\bf 33} 283001}    

\bibitem{Yan}
{Yang K, Li Z, K{\" o}nig J L K, R{\o}dland L, St{\aa}lhammar M, and Bergholtz E J 2023
Homotopy, Symmetry, and Non-Hermitian Band Topology
arXiv:2309.14416 [cond-mat.mes-hall]}

\bibitem{Oku2}
Okuma N, Sato M 2023 Annu. Rev. Condens. Matter Phys. {\bf 14} 83


\bibitem{GKM1}
G\'omez-Ullate D, Kamran N, and Milson R 2009 J. Math. Anal. Appl. {\bf 359}  352
 
 \bibitem{GKM2}
G\'omez-Ullate D, Kamran N, and Milson R 2020 J. Approx. Theory {\bf 162} 987

\bibitem{GKM3}
G\'omez-Ullate D, Kamran N and Milson R 2012 J. Math. Anal. Appl. {\bf 387} 410

\bibitem{Que1}
Quesne C 2008 J. Phys. {\bf A41} 392001

\bibitem{Que2}
Bagchi B, Quesne C, and Roychoudhury R  2009 Pramana J. Phys. {\bf A73} 337

\bibitem{Que3}
Quesne C 2009 SIGMA {\bf 5} 084

\bibitem{OS1}
Odake S and Sasaki R 2009 Phys. Lett. {\bf B679} 414

\bibitem{OS2}
Odake S and Sasaki R 2010 Phys. Lett. {\bf B684} 173

\bibitem{Roy1}
Midya B and Roy B 2009 Phys. Lett.  {\bf A373} 4117

\bibitem{Roy2}
Dutta D and Roy P 2010  J. Math. Phys. {\bf 51} 042101

\bibitem{Ho7}
Ho C-L 2011 , Ann. Phys. {\bf 326} 797

\bibitem{HOS}
Ho C-L,  Odake S, and Sasaki R 2011  SIGMA {\bf 7} 107

\bibitem{OS3}
Odake S and Sasaki R 2011 Physics Letters {\bf B702} 164

\bibitem{KMR}
{Koelink E, Morey L, and Rom{\' a}n P 2023
Matrix exceptional Laguerre polynomials
arXiv:2306.03223 [math.CA]}

\bibitem{BY}
{Banerjee S and Yadav R K 2023
One continuous parameter family of Dirac Lorentz scalar potentials associated with 
exceptional orthogonal polynomials
arXiv:2309.12965 [quant-ph]}

\bibitem{Que4}
Quesne C 2024 J. Math. Phys. 65 043501

\bibitem{Gran1} 
Grandati Y 2011 Ann. Phys. {\bf 326} 2074


\bibitem{Gran2} 
Grandati Y 2011  J. Math. Phys. {\bf  52} 103505

\bibitem{SUSY}  
Cooper F, Khare A, and  Sukhatme U 1995 Phys. Rep. {\bf 251} 267


\end{thebibliography}
\end{document}